\documentstyle[twocolumn,aps,]{revtex}
\begin{document}
\def\be{\begin{equation}}
\def\ee{\end{equation}}
\def\bea{\begin{eqnarray}}
\def\eea{\end{eqnarray}}
\renewcommand{\thefootnote}{\fnsymbol{footnote}}

\twocolumn[\hsize\textwidth\columnwidth\hsize\csname@twocolumnfalse%
\endcsname

\title{Fixed point analysis of a scalar theory with an external field}
\author{Alfio Bonanno$^{(1)}$ and Dario Zappal\`a$^{(2)}$}
\address{
 $^{(1)}$Istituto di Astronomia, Universit\`a di Catania\\
Viale Andrea Doria 6, 95125 Catania, Italy\\
 $^{(2)}$Dipartimento di Fisica, Universit\`a di Catania, and
INFN, sezione di Catania\\
Corso Italia 57, 95128 Catania, Italy\\
}

\date{\today}
\maketitle
\draft
\begin{abstract}
A momentum dependent projection of the Wegner-Hougton 
equation is derived for a scalar theory coupled to an 
external field. This formalism is useful to discuss 
the phase diagram of the theory. In particular we study 
some properties of the
Gaussian fixed point.
\end{abstract} 

]
% \narrowtext

In field theory there are several physical situations where the scalar 
self-interacting theory  plays a role of paramount importance. 
In some cases it is useful to introduce the order parameters, 
lorentz-scalar from which one can extract informations on the 
ground state properties when macroscopic quantities like
the temperature or the density are changed. In QCD 
the representative examples are given by the two   
flavour singlet quantities namely the Polyakov loop and
the trace of the fermion condensate. The first  
signals the appearance of the deconfining transition and 
the second one the chiral symmetry restoration. 

On the other hand, the scalar theory has a central role 
in the Standar Model and  GUT's.
%and from this point of view it
%can be considered as a fundamental field of the Nature. 
Therefore it is of great interest to have a careful understanding of its
fixed point structure in order to determine the phase diagram and 
all possible relevant directions in the parameter space.  

In the framework of the
$\epsilon$-expansion for the scalar theory,
one finds only the Gaussian fixed point
in any dimensions and the Wilson-Fisher fixed point for $2< D< 4$ 
\cite{wilson}.
An interesting issue has been recently raised
in \cite{huang} where it is found that around the
Gaussian fixed point new relevant directions exist;
this result however is not widely accepted 
\cite{morrisprl}.

A convenient way to discuss these problems is provided by the 
Wegner-Houghton equation \cite{wegner}
which is obtained by performing a functional integration over the  
infinitesimal        
momentum shell $k, k-\delta k$ in the limit $\delta k \rightarrow 0$. 
The above functional equation, which in principle is exact and 
contains the whole effect due to the integration of the fast modes,
has no known exact analytical solution 
and one has to resort to approximations in order to deal with a more 
tractable problem.
In \cite{hasen} a momentum independent projection of the 
Wegner-Houghton equation has been used to obtain a renormalization
group equation for the local potential, and it has been used 
to classify the interactions around the Gaussian fixed point.
Later the $O(\partial^2)$ contributions have been included  
and the anomalous dimension has been calculated\cite{morris}.

In this Report we use the gradient expansion
 in order to project on 
a larger subset of the parameter space by including a generic 
field dependence in the wavefunction renormalization 
function\cite{fraser,polonio}.
We consider a system of  two coupled scalar fields $\phi$ and $\psi$,  
where the second one is treated as classical, i.e. its quantum
fluctuations are neglected. 
This framework is useful when one is interested in studying the
phase diagram for the field $\phi$ in the presence of an external 
field $\psi$, whose functional form is kept fixed, which in practice 
is realized  if the cut-off of the integrated field is much below the
mass of the other (external) field.
In fact another application is a problem
where a mass hyerarchy is present. For example in some cases one is
interested in integrating out a heavy field in order to construct
the effective theory in the low energy domain. In these cases it is important
to monitor the decoupling of the heavy field from the low energy
effective theory. Conversely one can discuss the effects on a heavy
field of the integration of an extremely light one. 
Here we concentrate on the first issue
and analyse how some properties of the fixed point structure of a scalar field
are modified by turning on a new scalar degree of freedom, which we treat as 
classical.

%In our case we generates the renormalization group transformation by 
%by calculating the blocked action $\tilde S$,
%\be
%exp(-{\tilde S}_{k-\delta k})= \int D[\tilde\phi]exp
%S_k(\Psi+\tilde\phi
%\ee 
We start with the euclidean action defined at the cut-off 
$k$  
\bea
&&S_{k}[\phi,\psi] = \int d^Dx \Big (\frac{1}{2}
Z_{k}(\phi,\psi)\partial_\mu\phi\partial^\mu\phi+ 
U_{k}(\phi,\psi)+
\nonumber\\
&&\frac{1}{2}W_{k}(\phi,\psi)\partial_\mu\psi\partial^\mu\psi+
Y_{k}(\phi,\psi)\partial_\mu\phi\partial^\mu\psi
\Big).
\label{start}                              
\eea
The renormalization group equation is obtained by means of the blocking
transformation 
applied to the $\phi$ field. By performing the functional 
integration in the path integral only on the components of the $\phi$ 
field with momentum within the shell $(k, k-\delta k)$,
we obtain  the blocked action
\bea\label{final}
&&{S}_{k-\delta k}[\Phi,\Psi]=
\int d^Dx \Big (\frac{1}{2}
{Z}_{k-\delta k}(\Phi,\Psi)\partial_\mu\Phi\partial^\mu\Phi+\nonumber\\
&&\frac{1}{2}W_{k-\delta k}(\Phi,\Psi)\partial_\mu\Psi\partial^\mu\Psi
+Y_{k-\delta k}(\Phi,\Psi)\partial_\mu\Phi\partial^\mu\Psi+\nonumber\\&&
U_{k-\delta k}(\Phi,\Psi)\Big)=\nonumber\\
&&S_{k}[\Phi,\Psi]+\frac{1}{2}
{\rm Tr}'{\rm ln} K-\frac{1}{2}(FK^{-1}F)
\eea
where 
\be
F=(\delta S_k/\delta\phi)|_{\phi=\Phi,\psi=\Psi}\;\;\;K=(\delta^2 S_{k}
/\delta\phi\delta\phi)|_{\phi=\Phi,\psi=\Psi} 
\ee
and the prime means that 
the trace is taken over the infinitesimal shell $(k,k-\delta k)$. We use the 
gradient expansion 
in order to calculate the trace in the
above expression. We set $\Phi(x)=\phi_0+\tilde{\phi}(x)$ and $\Psi(x)= \psi_0
+\tilde{\psi}(x)$ and we write $K=K_0+\delta K$ where 
$K_0=K(\phi_0,\psi_0)=Z_{k}(\phi_0,\psi_0)p^2+
\partial^2_{\phi}U_{k}(\phi_0,\psi_0)$, the momentum $p$ lies in
the shell, and $\delta K$ is quadratic in both $\tilde\phi$ and $\tilde\psi$.
Then, the following expansion is used
\bea\label{expa}
&&{\rm Tr}'{\rm ln}K=\nonumber\\
&&{\rm Tr}'\Big ({\rm ln}K_0+K_0^{-1}\delta K -\frac{1}{2}
K_0^{-1}\delta K K_0^{-1}\delta K+O(\delta K ^3) \Big)
\eea
and we compute the trace in the last two terms by retaining at the same time
operators in the momentum and coordinate representation, as long as
all the operators in one representation are on the right side of the operators
in the other representation. In the last term of the expression this order
is not fulfilled and we use commutation rules to obtain the right ordering. All 
required commutators can be deduced from the simple rule $[f(x),p_\mu]=
-i\partial_\mu f(x)$. Once the $p-x$ dependence  has 
been disentangled, one can perform the trace operation and identify in the
right hand side of Eq.(\ref{expa}) contributions to the local blocked
potential $U_{k-\delta k}$ and to the wavefunction 
renormalization functions $Z_{k-\delta k},W_{k-\delta k},Y_{k-\delta k}$. 
Had we
started with the action 
\bea
&&S_{k}[\phi,\psi] = \int d^Dx \Big (-\frac{1}{2}
z_{k}(\phi,\psi)\phi\Box\phi-\nonumber\\
&&\frac{1}{2}w_{k}(\phi,\psi)\psi\Box\psi
+y(\phi,\psi)\partial_\mu\phi\partial^\mu\psi+U(\phi,\psi)
\Big)
\label{start2}                              
\eea
the functions $z,w,y$ could have been expressed 
in terms of $Z,W,Y$, previously introduced,
through $Z=\partial_\phi(\phi z)$, $W=\partial_\psi(\psi w)$ and 
$Y=y+[\phi\partial_\psi z+\psi\partial_\phi w]/2$, so there is no
ambiguity in the method. 

After a rather long but straightforward calculation
one finds the following flow equations for the dimensionless
renormalized blocked potential $V=U k^{-D}$
(the subscript ${k}$ will be omitted from now on and we define 
$t=ln(k/\Lambda)<0$ where $\Lambda$ is a fixed UV scale)
\bea\label{renopot} 
&&{\partial V\over \partial t}=
{(D-2)\over 2}(x\partial_x V+y\partial_y V)-DV\nonumber\\
&&-a_D{\ln}({{\cal A}\over {\cal A}|_{x=y=0}})
\eea
and the wavefunction renormalization functions
\bea\label{renozet}
&&{\partial Z\over \partial t}=
{(D-2)\over 2}(x\partial_x Z+y\partial_y Z)-
a_D{\cal A}^{-1}_t \Bigl( \partial_x^2 Z-\nonumber\\
&&2{\cal A}^{-1}\partial_x Z
\partial_x {\cal A}+{\cal A}^{-2}Z\partial_x{\cal A}[\partial_x {\cal A}+
\partial_x Z]-\nonumber\\
&&{\cal A}^{-3}Z^2(\partial_x {\cal A})^2
\Bigr)\\[2mm]            
&&{\partial Y\over\partial t}=
{(D-2)\over 2}(x\partial_x Y+y\partial_y Y)
-a_D{\cal A}^{-1}\Bigl( \partial_x^2 Y-\nonumber\\
&&{\cal A}^{-1}(\partial_x Z \partial_y {\cal A}+
\partial_x Y \partial_x {\cal A})
+{\cal A}^{-2} Z
(\partial_x{\cal A}\partial_y {\cal A}+\nonumber\\
&&\partial_y {\cal A}\partial_x Z/8+\partial_x {\cal A}
\partial_y Z/8)-{\cal A}^{-3}Z^2\partial_x{\cal A}
\partial_y{\cal A}/8\Bigr)\\[2mm]  
&&{\partial W\over \partial t}={(D-2)\over 2}(x\partial_x W+y\partial_y W)
-a_D{\cal A}\Bigl( \partial_x^2 W-\nonumber\\
&&2{\cal A}^{-1}\partial_x Y
\partial_y {\cal A}+Z{\cal A}^{-2}
\partial_y {\cal A}(\partial_y{\cal A}+\partial_y Z)-\nonumber\\
&&Z^2{\cal A}^{-3}
(\partial_y {\cal A})^2 \Bigr )
\eea
where ${\cal A}(t,x,y)=Z(t,x,y)+\partial_x^2 V(t,x,y)$   
$a_D=1/2^D\pi^{D/2}\Gamma(D/2)$ and we have introduced the dimensionless
variables $x=\Phi k^{-(D-2)/2}, y=\Psi k^{-(D-2)/2}$.
 
The above set of equations has been used to discuss the Decoupling Theorem (DT)
\cite{appelquist} within 
%the Exact Renormalization Group (ERG) 
this
approach in \cite{como}. In particular,
making use of a polynomial expansion of $V$ and $Z$ in the two fields,
it has been shown that no 
deviation from the DT appears in this context but around the heavy mass
threshold 
some small effect of non-locality is present in the wave function 
renormalization $Z$.

%By ``external field '' we mean a field whose proper fluctuations can be 
%neglected,
%so that it can be thought as a ``classical'' field. 
%This is the case if, for example, cut-off of the field $\Phi$ 
%is much below the mass of the field $\Psi$. However we shall consider the 
%fluctuations which arise because of the interaction with 
%the $\Phi$ field, therefore we shall find non-zero renormalization for the
%$W$ and $Y$ terms.   %%
%
%One point should be stressed. When an external field is present on
%the system, it is not clear what we mean for self-similarity, since
%the presence of another fixed scale does not allow, in general,
%the system to be scale invariant....

Here, since we do not expect large effects due to $W$ and $Y$,
we just consider, for the analysis of the fixed point structure,
the two coupled equations
(\ref{renopot}),(\ref{renozet}) in the case
$2 < D \leq 4$. 
We begin our investigation by perturbing a fixed point of the theory
with $y=0$, with functions that in general depend on 
$y$.

Let us first consider the Gaussian fixed point, solution of Eqs.
(\ref{renopot}),(\ref{renozet}),
\be\label{gaussian}
V^{*}=0, \;\;\;\; Z^*={\rm const.=Z_0}
\ee
%this fixed point is also present in the presence of a constant external field
%therefore it is interesting to discuss the classification of the interactions
%around this point. 
%This corresponde to the situation 
%$$
%{\partial g_n[k,\Psi]\over \partial \Psi}=0
%\eqno(c3)$$
%in (c2). 
%Let us study the stability of this fixed point in the 
%presence of the external field. 
We thus write $V=\delta V$ and $Z=Z_0+\delta Z$ where the perturbed 
quantities depend on the external field and linearize the flow equations.
From (\ref{renopot}),(\ref{renozet}) we obtain
\bea\label{pert1}
&&{\partial\delta V\over \partial t}=
{(D-2)\over 2}(x\partial_x \delta V+y\partial_y \delta V)-\nonumber\\
&&D\delta V
-\alpha(\delta Z+\partial_x^2 \delta V)+C
\\[2mm]
&&{\partial \delta Z\over \partial t}=
{(D-2)\over 2}(x\partial_x \delta Z+y\partial_y \delta Z)
-\alpha\partial_x^2 \delta Z
\label{pert2}
\eea
where $C=\alpha (\delta Z+\partial^2_x\delta V)|_{x=y=0}$
and $\alpha=a_D/Z_0$.
In order to
study the quantized-$\lambda$ behavior,
we assume for the perturbations the  general form 

\be\label{qua}
\delta V=e^{-\lambda t}h(x)g(y)\;\;\;\;\;\;
\delta Z=e^{-\lambda t}s(x)r(y),
\ee 
therefore, since 
$t$ is non-positive,  $\lambda>0,\;=0,\;<0$ correspond respectively
to relevant, marginal or irrelevant directions.
Thus one obtains the eigenvalue equations
\bea
&&{D-2\over 2}(g(y)x\partial_x h(x)+h(x)y\partial_y g(y))
-Dh(x)g(y)-\nonumber\\
&&\alpha(g(y)\partial_x^2 h(x)+s(x)r(y))+C_0=
-\lambda h(x)g(y)
\label{eigen}
\\[3mm]
&&{D-2\over 2}(r(y)x\partial_x s(x)+s(x)y\partial_y r(y))\nonumber\\
&&-\alpha r(y)\partial_x^2 s(x)=-\lambda s(x)r(y)
\label{eigenzet}
\eea
where $C_0=e^{\lambda t}C$.
The solution of these two coupled equations,
for any positive integer $m$, non-negative integer $n$ and arbitrary 
integration constant $C_2\neq 0$, is 
\bea
\label{sol1}
&&\delta V(x,y)=-{\alpha\over D}C_2 e^{-\lambda_{mn}t}H_n(\beta x) y^m
\\[2mm]
\label{sol2}
&&\delta Z(x,y)= C_2 e^{-\lambda_{mn}t} H_n(\beta x) y^m
\eea
where 
\be
\label{lam1}
\lambda_{mn}=-{m+n\over 2}(D-2)
\ee
and
$H_n(\beta x)$ are the Hermite polynomials and
$\beta=\sqrt{(D-2)/(4\alpha)}$.

Eq. (\ref{lam1}) shows that 
for the values chosen for 
$m$ and $n$,
all eigen-directions around the fixed point
are irrelevant. 

It must be remarked that, since we are in the presence of 
an external field,
we retain solutions with 
odd $m$ and $n$, corresponding  to unbounded potentials; 
obviously, when a specific problem is considered, one must reject 
all solutions that are unphysical for that particular case.

%Note that in principle, (\ref{sol1}) and (\ref{sol2})
%are still solution for $n$ positive odd integer, 
%but they can be rejected since the corresponding potential is unbounded
%(the unboundedness of $\delta V$ for $y \to\infty$ is here allowed since 
%we consider the potential as a function of $x$ keeping $y$ fixed).

Negative integer values of $m$ instead give bounded solutions 
that, however, are singular in $y=0$. We shall comment on these later.

A slightly different solution, involving an additive constant
in the potential $\delta V$, is obtained by selecting
$m=0$ and, again, any 
non-negative integer $n$ in the eigenvalue (\ref{lam1})

\bea 
\label{solm}
&&\delta V = -{\alpha \over D} C_2 e^{-\lambda_{0n} t} 
(H_n(\beta x)- H_n(0))\\[2mm]
&&\delta Z = C_2 e^{-\lambda_{0n} t} H_n(\beta x)
\eea

Obviously the latter case corresponds to a decoupling of the field $y$
which does not appear in the solutions, and again all
perturbations are irrelevant except if $n=0$ where the interaction
is instead marginal. However, since $H_0(\beta x)$ is a constant,
it is easy to realize the $n=0$ solution is nothing else than
the Gaussian fixed point (\ref{gaussian}), which explains the marginality
of the solution.

The set of solutions displayed so far is not complete since 
the two equations (\ref{pert1}), (\ref{pert2}) 
decouple for the particular choice 
$\delta Z=0$.
In this case the equation for $\delta V$ only has to be solved and this 
yields new  solutions which
read ($n$ is again any non-negative integer and $C_2$ an
arbitrary non-vanishing integration constant)
\be\label{sol3}
\delta V= e^{-\lambda_{mn}t} C_2 y^m H_n(\beta x), 
\ee
with positive integer $m$, and 
\be\label{sol4}
\delta V= e^{-\lambda_{0n}t} C_2 (H_n(\beta x)-H_n(0)) 
\ee 
with  $m=0$. This time $\lambda_{mn}$ is
\be
\lambda_{mn}=-{(m+n)\over 2}(D-2)+D 
\ee
which, for the latter case with $m=0$ gives the well known 
classification of the polynomial interactions when $y$ is turned off. 

Summarizing, 
any perturbation in the external field which is present in the 
wavefuntion renormalization function $\delta Z$,
also appears in the 
potential and provides only irrelevant directions in the parameter space 
with eigenfunctions that behave like 
$H_n y^m$ where $H_n$ is a rescaled Hermite polynomial. 
The case corresponding to a constant non-vanishing  
$\delta Z$ is marginal, but the solution is just the Gaussian fixed point.
Finally, for $\delta Z=0$,  
the equation for $\delta V$ decouples and we have
new marginal and relevant perturbations due the presence 
of the external field. In particular, we note that in $D=4$
the coupling corresponding to the mass of the external field is
a relevant interaction, although no integration on the external field modes
has been performed in the blocking procedure.  

In this analysis we have not included the solutions 
with negative integer values of $m$, that can correspond to positive 
$\lambda_{mn}$ in Eq. (\ref{lam1}) and thus to relevant directions,
but are singular in $y=0$.
They can be formally taken into account, provided one does not turn off the 
field $y$ and replaces the normalization introduced in 
Eq. (\ref{pert1}) through the constant $C$, 
which would be ill defined, with another one that is suitable for this case.
However the limit where the influence of
the external field becomes small and then disappears 
is now reached for $y\to \infty$, whereas for $y<<1$ the physics of the field 
$x$ is strongly modified by the other field $y$. One can think of 
couplings like $x/y$ in the lagrangian  that could play such a role,
but we do not see any clear physical meaning that can be attached 
to  these solutions.
 
The search for non-trivial fixed points is a difficult task due to 
the non-linearity of the RG equations. For $2< D <4$ one finds
the ferromagnetic fixed point with only one relevant direction.
The asymtpotic form of the potential for large field 
in this fixed point can be found in \cite{morris} and, in principle,
it is possible to study the modifications
of the eigen-directions in the presence of 
an external field, as we have done for the gaussian fixed point,
through a numerical investigation of  equations 
(\ref{renopot}), (\ref{renozet}). 

From a different point of view our equations can be used to study 
the $t$ invariant solutions of (\ref{renopot}), (\ref{renozet}) 
with the external field turned on.
In this case the external field acts on the system as a new scale. 
%change of location 
%of a given fixed point under the effect of the
%external field. 
%and it determines the phase diagram of the theory. 
In fact Eq. (\ref{renopot})
with the left hand side equal to zero, can be recast in the form 
\be
{\partial V^* \over \partial \Lambda} = {(D-2)\over 2}x\partial_x V^*
-DV^*-a_D{\rm \ln}({{\cal A}^*\over {\cal A}^*|_{x=y=0}})
\label{ventiq}
\ee
with $-\Lambda = (2/(D-2)){\rm ln}y$ where it is manifest the role played
by the external field. 
%One can then study the phase diagram by 
%approaching the critical system determined
%by the external field.
For instance, if a solution of Eq. (\ref{ventiq}) exists, 
it is determined by the strength of the external field 
and by three arbitrary constants $\alpha$, $\beta$ and $\gamma$. 
Its asymptotic form, for large $x$ is 
\be
\label{wil3}
Z^* = \alpha+{y^2 \over x^2}+
+O(x^{-8})
\ee

\bea
V^* = \beta x^6+y^2 x^4+\gamma y^4 x^2-
{a_D\over 3}{\rm ln}{30 \alpha\beta\over 
{\cal A}^*|_{x=y=0}}-\nonumber\\[2mm]
{4 a_D\over 3}{\rm ln}x-
{2 a_D/9}-
{a_D\over 150 \beta x^4}-{2 y^2\over 15 \beta x^2}+O(x^{-6})
\label{wil4}
\eea
and we have supposed $y\ll x$ and we have worked out the expressions 
up to $O(y^4/x^4)$ terms. 
%The location of the critical 
%bare mass now depends on $y$ and also on the value of the anomalous
%dimension. However 
An interesting question is whether or not one recovers the Wilson fixed point
in $y=0$ starting from a non-zero external field strength as in Eqs.
(\ref{wil3}),(\ref{wil4}).
If this is not the case one should question the uniticity of the solution 
of the fixed point equations in $y=0$.
A numerical investigation of the equations
is mandatory in order to answer this issue.
We hope to address this point in a following communication.

\acknowledgments
%\vspace{1 cm}

The authors would like to thank Janos Polonyi for his constant advice
for many enlightening discussions.

\end{document}